\documentclass[preprint,12pt]{elsarticle}

\usepackage{graphicx}
\usepackage{dcolumn}
\usepackage{bm}
\usepackage{amssymb}
\usepackage{graphicx}
\usepackage{hyperref}
\usepackage{color}

\begin{document}

\begin{frontmatter}

\title{Mathematical model of brain tumour growth with drug resistance}

\author{Jos\'e Trobia$^{1,2}$, Kun Tian$^3$, Antonio M Batista$^{1,2,4}$, Celso
Grebogi$^{3,5}$, Hai-Peng Ren$^{3,6}$, Moises S Santos$^4$, Paulo R
Protachevicz$^2$, Fernando S Borges$^7$, Jos\'e D Szezech Jr$^{1,2}$, Ricardo L
Viana$^8$, Iber\^e L Caldas$^4$, Kelly C Iarosz$^{4,9,*}$}
\address{$^1$Department of Mathematics and Statistics, State University of
Ponta Grossa, 84030-900, Ponta Grossa, PR, Brazil}
\address{$^2$Graduate in Science Program - Physics, State University of
Ponta Grossa, 84030-900, Ponta Grossa, PR, Brazil}
\address{$^3$Shaanxi Key Lab of Complex System Control and Intelligent
Information Processing, Xi'an University of Technology, Xi'an
710048, PR, China}
\address{$^4$Institute of Physics, University of S\~ao Paulo, 05508-900,
S\~ao Paulo, SP, Brazil}
\address{$^5$Institute for Complex Systems and Mathematical Biology,
University of Aberdeen, AB24 3UE, Aberdeen, Scotland, United Kingdom}
\address{$^6$Xi'an Technological University, Xi'an, 710021, PR, China}
\address{$^7$Center for Mathematics, Computation, and Cognition, Federal
University of ABC, 09606-045, S\~ao Bernardo do Campo, SP, Brazil}
\address{$^8$Department of Physics, Federal University of Paran\'a, 80060-000,
Curitiba, PR, Brazil}
\address{$^9$Faculdade de Tel\^emaco Borba, FATEB, 84266-010, Tel\^emaco Borba,
PR, Brazil}

\cortext[cor]{Corresponding author: kiarosz@gmail.com,
antoniomarcosbatista@gmail.com}

\date{\today}


\begin{abstract}  
Brain tumours are masses of abnormal cells that can grow in an uncontrolled way
in the brain. There are different types of malignant brain tumours. Gliomas are
malignant brain tumours that grow from glial cells and are identified as
astrocytoma, oligodendroglioma, and ependymoma. We study a mathematical model
that describes glia-neuron interaction, glioma, and chemotherapeutic agent. In
this work, we consider drug sensitive and resistant glioma cells. We show that
continuous and pulsed chemotherapy can kill glioma cells with a minimal loss of
neurons. 
\end{abstract}


\end{frontmatter}


\section{Introduction}

Tumour cells are abnormal cells that are classified into benign and malignant.
The benign tumours do not invade the normal tissue, while the malignant tumour
invade and can spread around the body \cite{cooper2000}. The malignant tumours
are cancerous tumours and they have a growth rate much faster than normal cells
\cite{perry2008, tubiana1989}. Cancer is one of the main causes of death
worldwide and many treatments have been developed, such as chemotherapy,
surgery, and radiation therapy \cite{karlsson16}.

{\color{red} Thermodynamic analysis related to the energy management of cancer
cells was developed by Lucia and Grisolia \cite{lucia18}. In 2013, Lucia
\cite{lucia13} proposed the use of engineering thermodynamic approach on data
collected from brain and breast tumours, as well as, therapy based on entropy
generation approach \cite{lucia15a}. One example of this type of therapy is the
application of electromagnetic fields \cite{lucia15b,lucia15c}. Recently,
Bergandi et al. \cite{bergandi19} showed the efect of very low frequency
electromagnetic field in the cancer growth and developed a thermodynamic model
to obtain the type of frequency.} Mathematical modelling of tumour growth has
been used to understand different aspects of cancer
\cite{lopez17,lopez19a,lopez19b}.Pinho et al. \cite{pinho02} analysed a
mathematical model of cancer treatment by chemotherapy agent taking metastasis
into account. Borges et al. \cite{borges14} used a model to study tumour growth
under treatment by continuous and pulsed chemotherapy. Nani and Freedman
\cite{nani00} studied cancer immunotherapy through models that incorporate
tumor-immune interaction \cite{pillis05}. 

One of the most common type of malignant brain tumour is the glioma that starts
in the glial cells \cite{stupp07}. The glial cells provide neuronal support
and protection \cite{jakel17}. In the literature, it is possible to find
different brain tumour models. Partial \cite{rockne09} and ordinary
\cite{kansal00a} differential equations have been used to simulate the dynamic
behaviour related to the glioma growth.

Drug resistance in cancer is a major problem in chemotherapy treatment
\cite{peters18}, due to the ability of cancerous cells to develop resistance to
chemotherapeutic agents \cite{housman14}. Nass and Efferth \cite{nass18}
studied drug targets and resistance mechanisms in myeloma. Recently, He et al.
\cite{he18} reported mechanisms related to drug-resistant ovarian-cancer cells.
A mathematical modelling of therapy, inducing cancer drug resistance, was
analysed by Sun et al. \cite{sun16}. We propose a model with glioma drug
resistance by adding a new differential equation in the model proposed by
Iarosz et al. \cite{iarosz15} for gliomas with glia-neuron interactions and
chemotherapy treatment. The authors computed the values of the infusion of
chemotherapy agents in which the glioma is suppressed and a minimum number of
neurons is lost, without neurogenesis. In this work, the novelty is that we
consider glioma drug resistance. In this way, our model has glia-neuron
interactions, resistant and sensitive gliomas, as well as chemotherapy
treatment. The tumour treatment occurs through continuous or pulsed
chemotherapy. In the continuous chemotherapy, the neuronal lifespan depends on
the infusion of chemothe\-rapy agent rate and the mutation rate from
drug-sensitive to drug-resistant cells. With regard to the pulsed chemotherapy,
we show that the chemotherapy cycle and the time interval of the drug
application play important roles in the treatment.

This paper is organised as follows: in Section 2 we introduce the mathematical
model, Section 3 presents our results for continuous and pulsed chemotherapy,
and in the last Section we draw our conclusions.


\section{Brain tumour model with drug resistance}

\begin{figure}[htb]
\begin{center}
\includegraphics[scale=0.7]{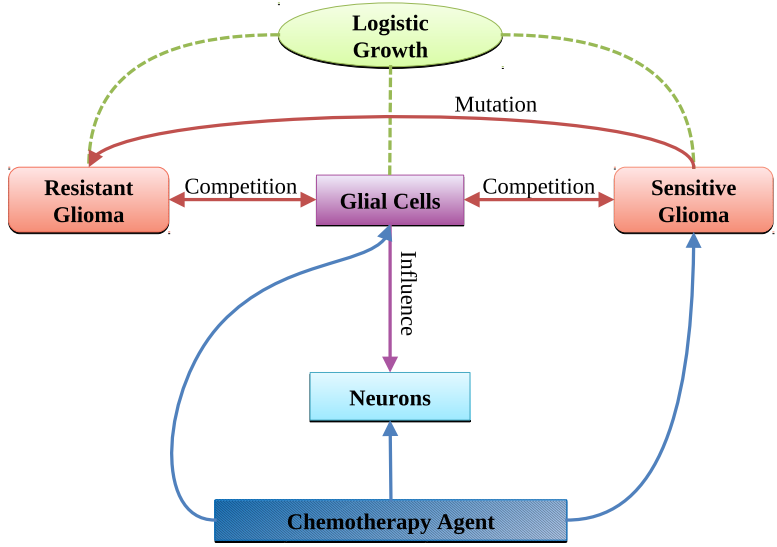}
\caption{(Colour online) Schematic representation of the model.}
\label{fig1}
\end{center}
\end{figure}

We include drug resistance in the Iarosz model \cite{iarosz15}. Figure
\ref{fig1} displays a schematic representation with the interactions considered
in the modified model. The sensitive and resistant glioma cells have logistic
growth, as well as they attack the glial cells and do not attack the neurons.
The glial cells interact with the neurons, attack the glioma cells, and have
logistic growth. The chemotherapy agent is a predator that attack the glioma
cells, glial cells, and neurons. Due to the chemotherapy, the sensitive glioma
cells convert to resistant glioma cells through mutations.

The mathematical model is described by
\begin{eqnarray}
\frac{dG(t)}{dt}&=& P_{G}G(t)\left(1-\frac{G(t)}{C_{1}}\right)-\Psi_{G}G(t)[S(t)
                    +R(t)]-\frac{I_1G(t)Q(t)}{A_1+G(t)},\\
\frac{dS(t)}{dt}&=& P_{S}S(t)\left(1-\frac{S(t)+R(t)}{C_2}\right)
                    -\Psi_{S}G(t)S(t)-uF[Q(t)]S(t) \nonumber \\
& & -\frac{I_{2}S(t)Q(t)}{A_{2}+S(t)}, \\
\frac{dR(t)}{dt}&=& P_{R}R(t)\left(1-\frac{S(t)+R(t)}{C_2}\right)
                    -\Psi_{R}G(t)R(t)+uF[Q(t)]S(t),\\
\frac{dN(t)}{dt}&=& \psi\dot{G}(t)F\left(-{\frac{\dot{G}(t)}{C_{1}}}\right)N(t)
                    -\frac{I_3N(t)Q(t)}{A_3+N(t)},\\
\frac{dQ(t)}{dt}&=& \Phi - \zeta Q(t),
\end{eqnarray}
where $G$ is the glial cells concentration (kg.m$^{-3}$), $S$ is the drug
sensitive glioma cells concentration (kg.m$^{-3}$), $R$ is the resistant drug
glioma cells concentration (kg.m$^{-3}$), $N$ is the neurons concentration
(kg.m$^{-3}$), $Q$ is the che\-motherapeutic agent concentration (mg.m$^{-2}$),
and $F(x)$ is a function defined as
\begin{equation}
F(x)=\left\{
\begin{array}{rcl}
0,  &\mbox{} & x\leq 0,\\  
1,  &\mbox{} & x>0.
\end{array}
\right. 
\end{equation}
Table \ref{tab1} describes the values of the parameters taken from the
referenced literature. In Eqs. (2) and (3), the third term is related to the
change from sensitive to resistant glioma cells. In Eq. (4), the first term
is associated with the decay of the neuronal population due to the glial cells
death.

The normalised model is given by
\begin{eqnarray}
\frac{dg(t)}{dt}&=&P_{G}g(t)[1-g(t)]-\beta_{1}g(t)[s(t)+r(t)]
                   -\frac{i_1g(t)Q(t)}{a_1+g(t)}, \\
\frac{ds(t)}{dt}&=&P_{S}s(t)[1-(s(t)+r(t))]-\beta_{2}g(t)s(t)-uF[Q(t)]s(t)
\nonumber \\
                & & -\frac{i_{2}s(t)Q(t)}{a_{2}+s(t)}, \\
\frac{dr(t)}{dt}&=&P_{R}r(t)[1-(s(t)+r(t))]-\beta_{3}g(t)r(t)+uF[Q(t)]s(t), \\
\frac{dn(t)}{dt}&=&\alpha\dot{g}(t)F[-\dot{g}(t)]n(t)-\frac{i_3n(t)Q(t)}{a_3
+n(t)}, \\
\frac{dQ(t)}{dt}&=&\Phi - \zeta Q(t),
\end{eqnarray}
where 
$g(t)=\frac{G(t)}{C_{1}}$, $s(t)=\frac{S(t)}{C_2}$, $r(t)=\frac{R}{C_2}$,
$n(t)=\frac{N(t)}{C_3}$, $\beta_{1}=\Psi_{G}C_{2}$, $\beta_{2}=\Psi_{S}C_{1}$,
$\beta_{3}=\Psi_{R}C_{1}$,  $\alpha=\psi C_{1}$, $a_{i}=\frac{A_{i}}{C_{i}}$,
and $i_{i}=\frac{I_{i}}{C_{i}}$ ($i=1,2,3$). The values of the parameters are
given in Table \ref{tab2}.

\begin{table}[htbp]
\centering
\caption{Parameters values taken from the referenced literature.}
\begin{tabular}{c|c|c} 
\hline
Parameter          & Values                         & Description\\ \hline
$P_{G}$            & $0.0068$ day$^{-1}$              & \\
$P_{S}$             & $0.012$ day$^{-1}$              &  Proliferation rate \cite{pinho13, spratt64} \\
$P_{R}$             & $0.002$ and $0.006$ day$^{-1}$   &  \\ \hline
$\psi$             & $0-0.02$                        & Loss influences \cite{pinho13}  \\ \hline
$I_1, I_3$          & $2.4\times 10^{-5}$ m$^{2}$(mg$\cdot$day)$^{-1}$ & Interaction\\
$I_2$               & $2.4\times 10^{-2}$ m$^{2}$(mg$\cdot$day)$^{-1}$ & coefficients \cite{pinho13, rzeski04} \\ \hline
$\Phi$              & $0-200$ mg(m$^{2}$.day)$^{-1}$   & Chemotherapy \cite{stupp05,strik12}   \\ \hline
$\zeta$             & $0.2$ day$^{-1}$                & Absorption rate \cite{borges14}   \\ \hline
$u$                 & $0-1$                          &   Mutation rate   \\ \hline
$A_1,A_2,A_3$        & $510$ kg.m$^{-3}$               & Holling type 2    \\ \hline
$\Psi_{G}$           & $3.6\times 10^{-5}$ day$^{-1}$   & Competition \\
$\Psi_{S},\Psi_{R}$   & $3.6\times 10^{-6}$ day$^{-1}$   &coefficients \cite{pinho13}\\ \hline
$C_1,C_2,C_3$        & $510$ kg.m$^{-3}$               & Carrying capacity \cite{azevedo09} \\ \hline
\end{tabular}
\label{tab1}
\end{table}

\begin{table}[htbp]
\centering
\caption{Values of the parameters for the normalisation.}
\begin{tabular}{c|c}
\hline 
Parameter            & Values \\ \hline 
$\beta_{1}$           & $1.8\times 10^{-2}$ day$^{-1}$                  \\ \hline
$\beta_{2},\beta_{3}$ & $1.8\times 10^{-3}$ day$^{-1}$                   \\ \hline
$\alpha$             & $0-10$                                         \\ \hline
$a_1,a_2,a_3$         & $1$                                            \\ \hline
$i_1,i_3$             & $4.7\times 10^{-8}$ m$^{2}$(mg$\cdot$day)$^{-1}$ \\ \hline
$i_2$                 & $4.7\times 10^{-5}$ m$^{2}$(mg$\cdot$day)$^{-1}$ \\ \hline
\end{tabular}
\label{tab2}
\end{table}

The equilibria points, which are physiologically feasible,
$E(\overline{g},\overline{s},\overline{r},\overline{n},\overline{Q})$ of the
model are obtained through $\dot{g}(t)=0$, $\dot{s}(t)=0$, $\dot{r}(t)=0$,
$\dot{n}(t)=0$, and $\dot{Q}(t)=0$. First, we analyse the local stability for
an undesirable equilibrium, where this equilibrium is given by
$E_0(0,0,0,0,\Phi { \zeta}^{-1})$. The eigenvalues of the Jacobian matrix are
\begin{eqnarray}
\lambda_1^{(0)}&=&P_G-\frac{i_1 \Phi}{\zeta a_1},\\
\lambda_2^{(0)}&=&P_S-\frac{i_2 \Phi}{\zeta a_2} - u,\\
\lambda_3^{(0)}&=&P_R,\\
\lambda_4^{(0)}&=&-\frac{i_3\Phi}{\zeta a_3},\\
\lambda_5^{(0)}&=&-\zeta.
\end{eqnarray}
We identify the stability of the equilibrium through the sign of the real part
of each eigenvalue. In a hyperbolic equilibrium, if the real part of each 
eigenvalue is strictly negative, then the equilibrium is locally asymptotically
stable, and if positive, then the equilibrium is unstable. In order to ensure
the stability of $E_0(0,0,0,\Phi{\zeta}^{-1})$, it is necessary that
\begin{eqnarray}
\Phi & > & \frac{P_G a_1 \zeta}{i_1}, \\
\Phi & > & \frac{(P_S - u) a_2 \zeta}{i_2},
\end{eqnarray}
where these results are obtained by means of $\lambda_1^{(0)}<0$ and 
$\lambda_2^{(0)}<0$. The values of the normalised parameters are positive 
(table \ref{tab2}), then the eigenvalues $\lambda_4^{(0)}$ and $\lambda_5^{(0)}$ 
are negative. However, the eigenvalue $\lambda_3^{(0)}$ is positive. Therefore,
the equilibrium $E_0(0,0,0,0,\Phi { \zeta}^{-1})$ is unstable, due to the fact
that the resistant drug glioma cells are not affected by the chemotherapeutic
agent. It is possible to find a stable equilibrium
$E_1(0,0,r^*,0,\Phi { \zeta}^{-1})$ for $r^*=1$. The eigenvalues of the Jacobian
matrix are
\begin{eqnarray}
\lambda_1^{(1)}&=&P_G-\beta_1-\frac{i_1 \Phi}{\zeta a_1},\\
\lambda_2^{(1)}&=&-\frac{i_2 \Phi}{\zeta a_2} - u,\\
\lambda_3^{(1)}&=& - P_R,\\
\lambda_4^{(1)}&=&-\frac{i_3\Phi}{\zeta a_3},\\
\lambda_5^{(1)}&=&-\zeta.
\end{eqnarray}
In order to ensure the stability of $E_1(0,0,r^*,0,\Phi{\zeta}^{-1})$, it is
necessary that
\begin{equation}
\Phi>\frac{(P_G -\beta_1) a_1 \zeta}{i_1 },
\end{equation}
where these results are obtained through $\lambda_1^{(0)}<0$. The eigenvalues
$\lambda_2^{(0)}$, $\lambda_3^{(0)}$, $\lambda_4^{(0)}$, and $\lambda_5^{(0)}$ are
negative because the values of the normalised parameters are positive. We
consider $a_1=1$, $P_G=0.0068$, $\beta_1=0.018$, $i_1=4.7\times10^{-8}$, and
$\zeta=0.2$ (table \ref{tab2}). With these values, we obtain that $E_1$ is
linearly asymptotically stable for $\Phi>-47659,57$. Therefore, when the
chemotherapeutic agent kills all glial cells ($g$) and drug sensitive glioma
cells ($s$), the normalised resistant drug glioma cells concentration is $r=1$.

We also consider the equilibrium
$E_2(\overline{g},0,0,\overline{n},\overline{Q})$ that represents the complete
elimination of drug sensitive glioma cells and resistant drug 
glioma cells, in addition, the glial and neuron cells are preserved. This
equilibrium is obtained by the solution of
\begin{eqnarray}
P_G (1-\overline{g})-\frac{i_1 \overline{Q}}{a_1+\overline{g}} = 0,
\label{equib1}\\
-\frac{i_3 \overline{n} \overline{Q}}{a_3+\overline{n}}=0, \\
\Phi-\zeta\overline{Q}=0, \label{equib2}
\end{eqnarray}
for $\overline{n}=0$ and $\overline{Q}=\Phi{\zeta}^{-1}$. Thus, the 
equilibrium $E_2(\overline{g},0,0,\overline{n},\overline{Q})$ is given by 
$E_2(\overline{g},0,0,0,\Phi {\zeta}^{-1})$, meaning that all neurons are also
eliminated. Equation (\ref{equib1}) can be rewritten as
\begin{equation}\label{eqg}
{\overline{g}}^2+\overline{g}(a_1-1)-a_1+\frac{i_1\Phi}{\zeta P_G}=0
\label{eqE2}.
\end{equation}
Using the parameters of Table \ref{tab1} and \ref{tab2}, $\overline{g}$ has a
real, positive and non null solution when $\Phi<28936.17$. The eigenvalues of
the Jacobian matrix for $E_2$ are
\begin{eqnarray}
\lambda_1^{(2)}&=& P_G (1-2 \overline{g})-\frac{i_1 a_1 \Phi}{\zeta
(a_1+\overline{g})^2},\label{eq1E2l1}\\
\lambda_2^{(2)}&=& P_S - \beta_2 \overline{g}- u -\frac{i_2 \Phi}{\zeta a_2},\\
\lambda_3^{(2)}&=& P_R -\beta_3 \overline{g},\\
\lambda_4^{(2)}&=&-\frac{i_3\Phi}{\zeta a_3},\\
\lambda_5^{(2)}&=&-\zeta.
\end{eqnarray}
For $a_1=1$, $\lambda_1^{(2)}$ is negative in Equation (\ref{eq1E2l1}) when 
\begin{eqnarray}
(1+\overline{g})^2(1-2 \overline{g}) < \frac{ i_1 \Phi}{\zeta P_G}.
\end{eqnarray}
From Equation (\ref{eqE2}), it is obtained
$\frac{i_1\Phi}{\zeta P_G}=1-\overline{g}^2$, consequently
\begin{eqnarray}
(1-\overline{g}^2)-2\overline{g}^2-2\overline{g}^3<(1-\overline{g}^2),
\end{eqnarray}
therefore $\lambda_1^{(2)}<0$ if $\overline{g}>0$. $\lambda_2^{(2)}$ is negative
for combinations of $u$ and $\Phi$, for example: i) $u=0$ and $\Phi>43.41$, ii)
$u=0.001$ and $\Phi>39.15$, and iii) $u=0.01$ and $\Phi>0.85$. The values of 
$\lambda_4^{(2)}$ and $\lambda_5^{(2)}$ are negative. However, $\lambda_3^{(0)}$ 
is positive if $P_R>\beta_3$. Using the parameters from Tables \ref{tab1} and
\ref{tab2}, we obtain $\beta_3=0.0018$ and $P_R\ge 0.002$. For realistic
parameters, the equilibrium $E_2$ is unstable, due to the fact that the
proliferation rate of the resistant drug glioma cells is larger than the
normalized competition coefficient between glioma and resistant drug glioma
cells. 

The equilibrium $E_2$, which is related to the elimination of all glioma cells,
is unstable. In this case, all neurons are also eliminated, showing that it is
impossible find a cure for glioma in the drug resistant case. For this reason,
we focus on the neuronal lifespan during the chemotherapeutic treatment.


\section{Chemotherapy treatment}

\subsection{Continuous chemotherapy}

In continuous chemotherapy, the anticancer drugs are administered without
pauses \cite{vogelzang84}. Continuous infusion, followed by radiotherapy, was
used as a treatment for malignant tumour. Many researchers reported that this
combination can improve the tumours regression \cite{rotman91, lai17}. We
consider continuous chemotherapy as a way to eliminate glioma cells.

\begin{figure}[htb]
\begin{center}
\includegraphics[scale=0.5]{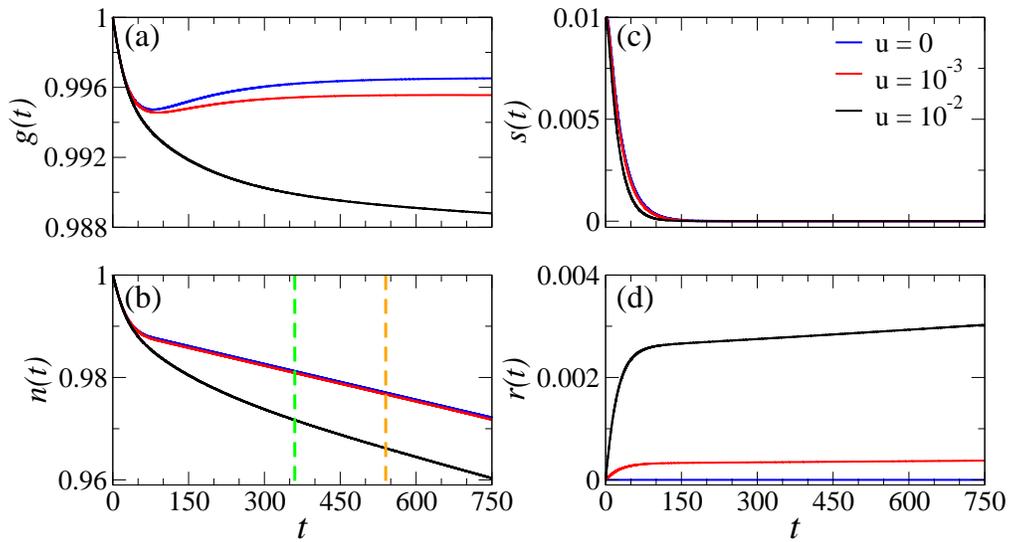}
\caption{(Colour online) Time evolution of (a) glial cells concentration
$g(t)$, (b) neurons concentration $n(t)$, (c) drug sensitive glioma cells
concentration $s(t)$, and (d) resistant drug glioma cells concentration $r(t)$
for $\Phi=200$, $P_R=0.002$, and there mutation rate $u=0$ (blue line),
$u=10^{-3}$ (red line), and $u=10^{-2}$ (black line). The green and orange
vertical dashed lines correspond to $360$ days ($12$ months) and $540$ days
($18$ months), respectively. We consider $g(0)=1$, $n(0)=1$, $s(0)=0.01$,
$r(0)=0$, and $Q(0)=0$.}
\label{fig2}
\end{center}
\end{figure}

Figure \ref{fig2} shows the time evolution of normalised (a) glial cells
concentration $g(t)$, (b) neurons concentration $n(t)$, (c) drug sensitive
glioma cells concentration $s(t)$, and (d) resistant drug glioma cells
concentration $r(t)$. We consider $\Phi=200$, $P_R=0.002$, $u=0$ (blue line),
$u=10^{-3}$ (red line), and $u=10^{-2}$ (black line). The chemotherapeutic agent
kills the glial cells, neurons, and sensitive glioma cells. For $u=0$, there is
not resistant glioma, and as a consequence the malignant tumour is suppressed.
However, $n$ decreases from $1$ to $0.981$ for $t=360$ days ($12$ months)
(green vertical dashed line) and to $0.977$ for $t=540$ days ($18$ months)
(orange vertical dashed line). The glial cells are killed by glioma and
chemotherapy, but they exhibit logistic growth and saturation. For $u=10^{-3}$,
$n$ goes to $0.980$ and $0.976$ for $360$ and $540$ days, respectively.
Considering $u=10^{-2}$, we observe $n=0.971$ for $t=360$ days, and $n=0.966$
for $t=540$ days. In addition, the sensitive glioma cells are absent for $t$
greater than approximately $150$ days.

\begin{figure}[htb]
\begin{center}
\includegraphics[scale=0.35]{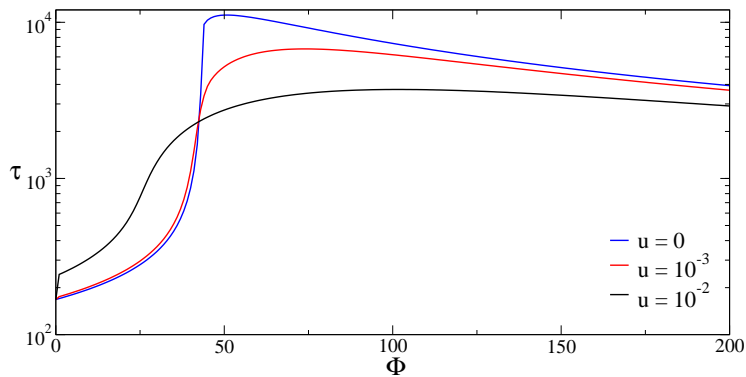}
\caption{(Colour online) $\tau$ as a function of $\Phi$ for $P_R=0.002$, $u=0$
(blue line), $u=10^{-3}$ (red line), and $u=10^{-2}$ (black line).}
\label{fig3}
\end{center}
\end{figure}

In Fig. \ref{fig3}, we see the time $\tau$ to achieve a neuron concentration
$n=0.9$ as a function of $\Phi$. For $u=0$ (blue line) and $\Phi=200$, $\tau$
is equal to $3926$ days, while $\tau$ is much less for $u=10^{-3}$ (red line)
and $u=10^{-2}$ (black line). We obtain $\tau$ equal to $3672$ and $2910$ days
for $u$ equal to $10^{-3}$ and $10^{-2}$, respectively. Then, it is possible to
verify that $u$ has an important effect on $\tau$.

\begin{figure}[htb]
\begin{center}
\includegraphics[scale=1]{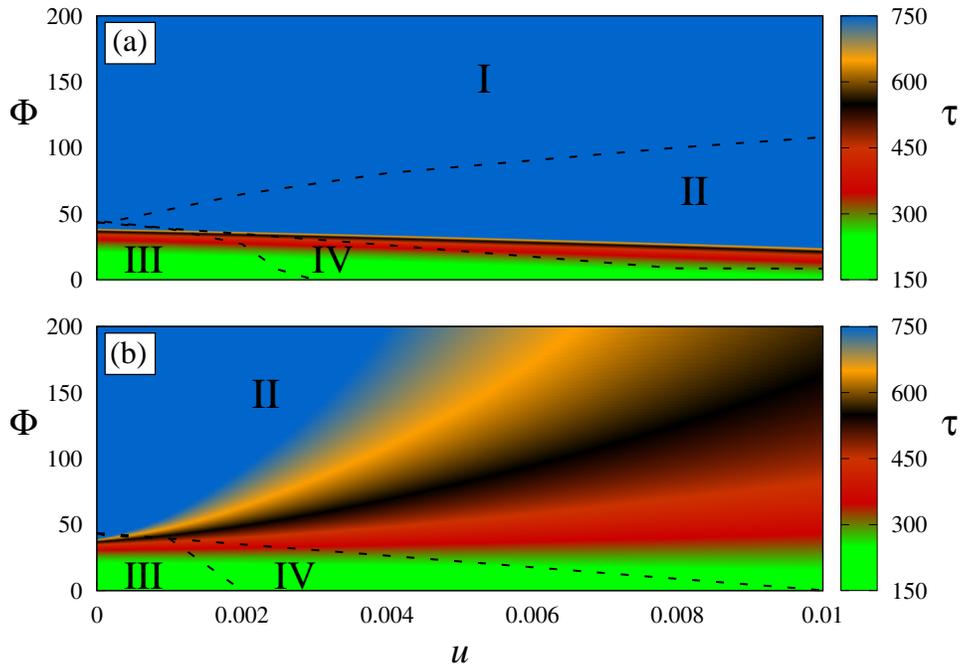}
\caption{(Colour online) Time ($\tau$) to achieve the concentration ($n=0.9$)
(colour bar) as a function of $\Phi\times u$ for (a) $P_R=0.002$ and (b)
$P_R=0.006$.}
\label{fig4}
\end{center}
\end{figure}

We also calculate $\tau$ by varying $\Phi$ and $u$, as shown in Fig.
\ref{fig4}. The colour bar represents the values of $\tau$. In our simulations,
the blue region corresponds to $\tau$ greater than $700$ days. In the orange,
black and red regions, the $\tau$ values are for about $650$, $500$ and $350$
days, respectively. The $\tau$ values less than $300$ days are in the green
region. Figure \ref{fig4}(a), for $P_R=0.002$, is separated into four regions
denoted by I, II, III, and IV. In the region I, we have $s(t)<0.01$ and
$r(t)<0.01$, namely when $n=0.9$ both sensitive and resistant gliomas have a
concentration less than the initial glioma concentration. Region II,
$s(t)<0.01$ and $r(t)>0.01$, shows that only the sensitive glioma is
suppressed. The sensitive glioma grows in region III, $s(t)>0.01$ and
$r(t)<0.01$. In the region IV, both sensitive and resistant glioma have a
concentration greater than the initial glioma concentration, $s(t)>0.01$ and
$r(t)>0.01$. For $P_R=0.006$ and $\Phi\leq 200$ (Fig. \ref{fig4}(b)), there is
no region I.

\subsection{Pulsed chemotherapy}

\begin{figure}[htb]
\begin{center}
\includegraphics[scale=0.45]{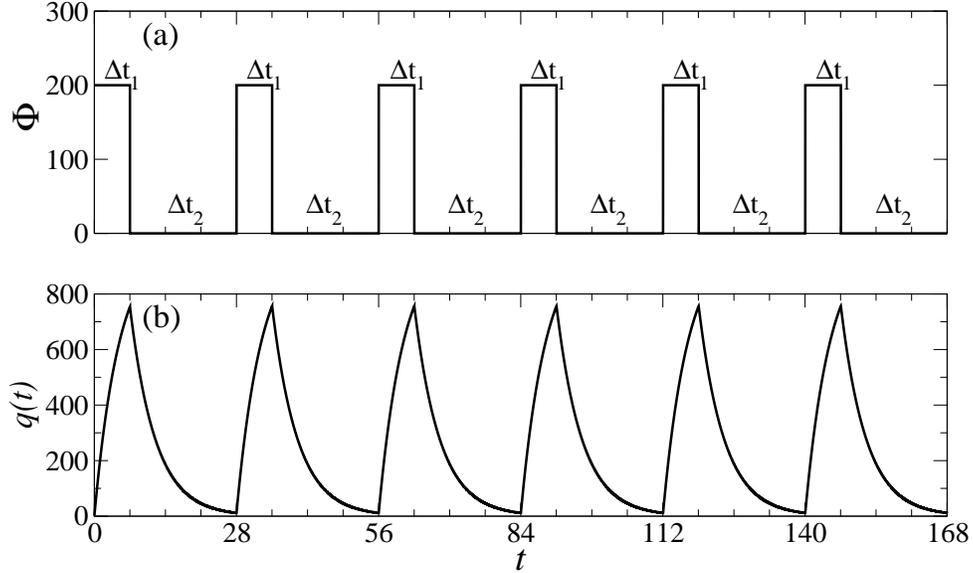}
\caption{(a) Intermittent schedule of the pulsed chemotherapy, where
$\Delta t_1$ and $\Delta t_2$ are the time intervals with (days on) and without
(days off) chemotherapy, respectively. (b) Temporal evolution of drug
concentration $q(t)$.}
\label{fig5}
\end{center}
\end{figure}

Pulsed chemotherapy is the use of intermittent schedules of chemotherapeutic
agents to treat diseases \cite{beer04}. Researchers have been carrying out
various treatment types with different protocols to eliminate cancerous cells.
In literature, it is possible to find results based on theoretical studies
\cite{foo09,ren17} and experiments \cite{wasan14}. Our intermittent schedule
is illustrated in Fig. \ref{fig5}(a), where $\Delta t_1$ and $\Delta t_2$
correspond to the time intervals with (days on) and without (days off)
chemotherapy, respectively. Figure \ref{fig5}(b) displays the temporal evolution
of $q(t)$. We observe an exponential growth of drug concentration $q(t)$ during
the drug application and an exponential decay after the application.

\begin{figure}[htb]
\begin{center}
\includegraphics[scale=0.45]{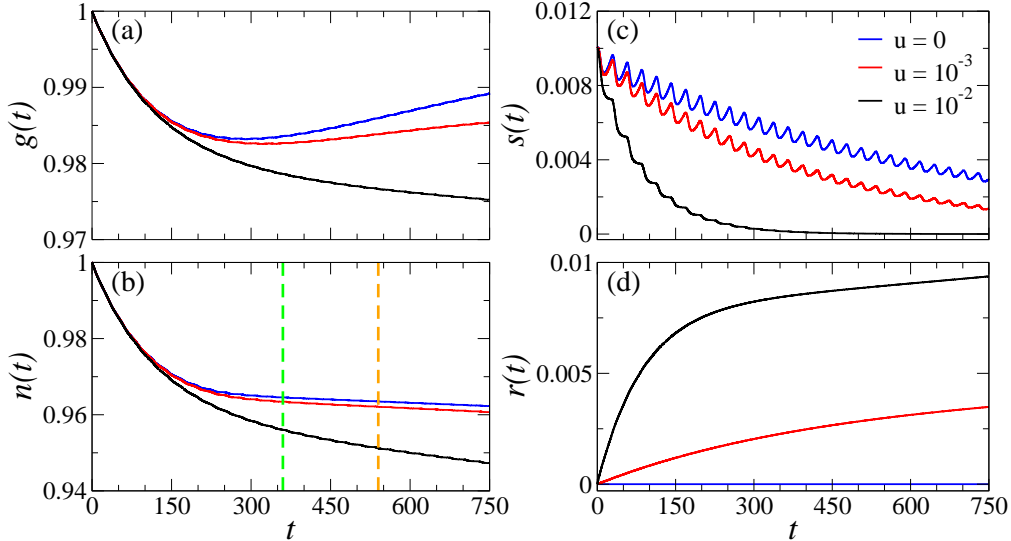}
\caption{(Colour online) Time evolution of (a) $g(t)$, (b) $n(t)$, (c) $s(t)$,
and (d) $r(t)$ for $\Delta t_1=7$ days with $\Phi=200$ and $\Delta t_2=21$ days
with $\Phi=0$, $P_R=0.002$, $u=0$ (blue line), $u=10^{-3}$ (red line), and
$u=10^{-2}$ (black line). The green and orange vertical dashed lines correspond
to $360$ and $540$ days, respectively.}
\label{fig6}
\end{center}
\end{figure}

Figure \ref{fig6} shows (a) $g(t)$, (b) $n(t)$, (c) $s(t)$, and (d) $r(t)$ for
$\Delta t_1=7$ days with $\Phi=200$ and $\Delta t_2=21$ days. We consider
$P_R=0.002$, $u=0$ (blue line), $u=10^{-3}$ (red line), and $u=10^{-2}$ (black
line). At $t=360$ days (green vertical dashed line), we find $n=0.964$,
$n=0.963$, and $n=0.956$ for $u=0$, $u=10^{-3}$, and $u=10^{-2}$ respectively.
When $t=540$ days, $n=0.963$ for $u=0$, $n=0.962$ for $u=10^{-2}$, and
$n=0.951$ for $u=10^{-2}$. In this intermittent schedule, the $n$ values are
less than the results found considering the continuous chemotherapy. 

\begin{figure}[htb]
\begin{center}
\includegraphics[scale=1]{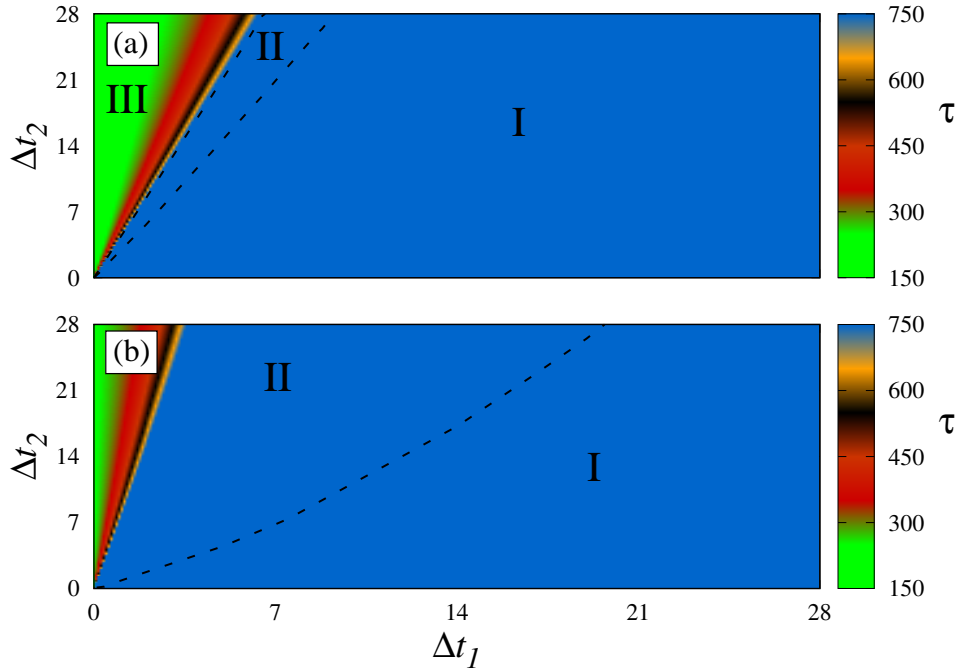}
\caption{(Colour online) $\tau$ (colour bar) as a function of
$\Delta t_2\times \Delta t_1$ for $P_R=0.002$, (a) $u=10^{-3}$ and (b)
$u=10^{-2}$.}
\label{fig7}
\end{center}
\end{figure}

There are many types of treatment schedules. With this in mind, we vary the
number of days on and off to analyse the effects of the drug resistance on
different chemotherapy protocols. Figure \ref{fig7} exhibits $\tau$ (colour
bar) as a function of $\Delta t_2\times \Delta t_1$. In Fig. \ref{fig7}(a), we
verify the existence of the four regions, where the region IV is very small and
it is between the regions II and III. For $u=10^{-2}$, there are only the
regions I and II, as shown in Fig. \ref{fig7}(b). The region I is larger for
$u=10^{-3}$ (Fig. \ref{fig7}(a)) than for $u=10^{-2}$ (Fig. \ref{fig7}(b)).
Therefore, the region I decreases and the region II increases when $u$
increases, i.e., the number of treatment schedules that control the growth of
both sensitive and resistant glioma cells decreases.

\section{Conclusions}

There are many different types of brain tumours. The treatments depend on the
tumour characteristics. One of the most common malignant tumours in the brain
is the glioma. This tumour begins in the glial cells and affects the support
for the neurons. Due to this fact, without support and protection, the number
of neurons decreases.

We extend the mathematical model of brain tumour growth proposed by Iarosz et
al. \cite{iarosz15}. The Iarosz model describes glia-neuron interaction and
chemotherapy treatment. In this work, we modify the model separating the
equation of the glioma cells into two equations. The new equations correspond
to sensitive and resistant glioma cells.

We consider continuous and pulsed chemotherapy to destroy glioma cells without
harming a large number of neurons. With regard to the continuous chemotherapy,
the time $\tau$ to achieve $n=0.9$ decreases when the mutation rate $u$ from
sensitive to resistant glioma cells increases. The $\tau$ values depend on
$\Phi$ and $u$. For small $P_R$ values, we find values in the parameter space
$\Phi\times u$ (region I) where the continuous chemotherapy kills both
sensitive and resitant gliomas. In the pulsed chemotherapy, the region of the
best treatment according to days on and off decreases for larger $u$ values.


\section*{Acknowledgements}
We wish to acknowledge the support: Arauc\'aria Fundation, National Council for
Scientific and Technological Development (CNPq), Coordination for the
Improvement of Higher Education Personnel (CAPES), and S\~ao Paulo Research
Foundation (Processes 2015/07311-7, 2017/18977-1, and 2018/03211-6). The
authors would like to thank the 105 Group Science (www.105groupscience.com) for
the fruitful discussions.



\begin{thebibliography}{10}

\bibitem{cooper2000}
G.M. Cooper, The Development and Causes of Cancer. The Cell: A Molecular
Approach 2nd edition, Sunderland (MA), Sinauer Associates (2000). 
\bibitem{perry2008}
M.C. Perry, The chemotherapy sourcebook, Baltimore, Williams $\&$ Wilkins
(2008).
\bibitem{tubiana1989}
M. Tubiana, Tumor Cell Proliferation Kinetics and Tumor Growth Rate, Acta
Oncologica 28 (1980) 113-121.
\bibitem{karlsson16}
P. Karlsson, B.F. Cole, K.N. Price, R.D. Gelber, A.S. Coates, A. Goldhirsch,
M.  Castiglione, M. Colleoni, G. Gruber, Timing of radiation therapy and
chemotherapy after breast-conserving surgery for node-positive breast cancer:
long-term results from international breast cancer study group trial VI and
VII. International Journal of Radiation Oncology, Biology, Physsics 96 (2016)
273-279.
\bibitem{lucia18}
U. Lucia, G. Grisolia, Constructal law and ion transfer in normal and cancer
cells, Proceedings of the romanian academy, Series A 2018 (2018) 213-218.
\bibitem{lucia13}
U. Lucia, Thermodynamics and cancer stationary states, Physica A 392 (2013)
3648-3653.
\bibitem{lucia15a}
U. Lucia, Bioengineering thermodynamics: An engineering science for
thermodynamics of byosystems, International Journal of Thermodynamics 18 (2015)
254-265.
\bibitem{lucia15b}
U. Lucia, A. Ponzetto, T.S. Deisboeck, A thermodynamic approach to the
'mitosis/apoptosis' ratio in cancer, Physica A 436 (2015) 246-255.
\bibitem{lucia15c}
U. Lucia, Bioengineering thermodynamics of biological cells, Theoretical Biology
and Medical Modelling 12 (2015) 1-16.
\bibitem{bergandi19}
L. Bergandi, U. Lucia, G. Grisolia, R. Granata, I. Gesmundo, A. Ponzetto, E.
Paolucci, R. Borchiellini, E. Ghigo, F. Silvagno, The extemely low frequency
electromagnetic stimulation selective for cancer cells elicits growth arrest
through a metabolic shift, BBB - Molecular Cell Research 1866 (2019)
1389-1397.
\bibitem{lopez17}
A.G. L\'opez, K.C. Iarosz, A.M. Batista, J.M. Seoane, R.L. Viana, M.A.F.
Sanju\'an, The dose-dense principle in chemotherapy, Journal of Theoretical
Biology 430 (2017) 169-176.
\bibitem{lopez19a}
A.G. L\'opez, K.C. Iarosz, A.M. Batista, J.M. Seoane, R.L. Viana, M.A.F.
Sanju\'an, Nonlinear cancer chemotherapy: Modelling the Norton-Simon
hypothesis, Communications in Nonlinear Science and Numerical Simulation 70
(2019) 307-317.
\bibitem{lopez19b}
A.G. L\'opez, K.C. Iarosz, A.M. Batista, J.M. Seoane, R.L. Viana, M.A.F.
Sanju\'an, The role of dose density in combination cancer chemotherapy,
Communications in Nonlinear Science and Numerical Simulation 79 (2019) 104918.
\bibitem{pinho02}
S.T.R. Pinho, H.I. Freedman, F. Nani, A chemotherapy model for the treatment of
cancer with metastasis, Mathematical and Computer Modelling 36 (2002) 773-803.
\bibitem{borges14}
F.S. Borges, K.C. Iarosz, H.P. Ren, A.M. Batista, M.S. Baptista, R.L. Viana,
S.R. Lopes, C. Grebogi, Model for tumour growth with treatment by continuous
and pulsed chemotherapy, BioSystems 116 (2014) 43-48.
\bibitem{nani00}
F. Nani, H.I. Freedman, A mathematical model of cancer treatment by
immunotherapy, Mathematical Biosciences 163 (2000) 159-199.
\bibitem{pillis05}
L.G. de Pillis, A.E. Radunskaya, C.L. Wiseman, A validated mathematical model
of cell-mediated immune response to tumor growth, Cancer Research 65 (2005)
7950-7958.
\bibitem{stupp07}
R. Stupp, M.E. Hegi, M.R. Gilbert, A. Chakravarti, Chemoradiotherapy in
malignant glioma: Standard of care and future directions, Journal of Clinical
Oncology 25 (2007) 4127-4136.
\bibitem{jakel17}
S. J\"akel, L. Dimou, Glial cells and their function in the adult brain: A
journey through the history of their ablation, Frontiers in Cellular
Neuroscience 11 (2017) 1-17.
\bibitem{rockne09}
R. Rockne, E.C. Alvord Jr., J.K. Rockhill, K.R. Swanson, A mathematical model
for brain tumor response to radiation therapy, Journal of Mathematical Biology
58 (2009) 561-578.
\bibitem{kansal00a}
Kansal AR, Torquato S, Chiocca EA, Deisboeck TS. Emergence of a subpopulation
in a computational model of tumor growth. J Theor Biol 2000;207:431-441.
\bibitem{peters18}
G.J. Peters, Cancer drug resistance: a new perspective, Cancer drug resistance 1
(2018) 1-5.
\bibitem{housman14}
G. Housman, S. Byler, S. Heerboth, K. Lapinska, M. Longacre, N. Snyder, S.
Sarkar, Drug resistance in cancer: An overview, Cancers 6 (2014) 1769-1792.
\bibitem{nass18}
J. Nass, T. Efferth, Drug targets and resistance mechanisms in multiple
myeloma, Cancer Drug Resistance 1 (2018) 87-117.
\bibitem{he18}
Y.J. He, K. Meghani, M.-C. Caron, C. Yang, D.A. Ronato, J. Bian, A. Sharma,
J. Moore, J. Niraj, A. Detappe, J.G. Doench, G. Legube, D.E. Root, A.D.
D'Andrea, P. Dran\'e, S. De, P.A. Konstantinopoulos, J.-Y. Masson, D.
Chowdhury, DYNLL1 binds to MRE11 to limit DNA end resection in BRCA1-deficient
cells, Nature 563 (2018) 522-526.
\bibitem{sun16}
X. Sun, J. Bao, Y. Shao, Mathematical modeling of therapy-induced cancer drug
resistance: Connecting cancer mechanisms to population survival rates.
Scientific Reports 6 (2016) 22498.
\bibitem{iarosz15}
K.C. Iarosz, F.S. Borges, A.M. Batista, M.S. Baptista, R.A.N. Siqueira, R.L.
Viana, S.R. Lopes,  Mathematical model of brain tumour with glia-neuron
interactions and chemotherapy treatment, Journal of Theoretical Biology 368
(2015) 113-121.
\bibitem{pinho13}
S.T.R. Pinho, F.S. Barcelar, R.F.S. Andrade, H.I. Freedman, A mathematical
model for the effect of anti-angiogenic therapy in the treatment of cancer
tumours by chemotherapy, Nonlinear Analysis: Real World Applications 14 (2013)
815-828.
\bibitem{spratt64}
J.S. Spratt, T.L. Spratt, Rates of growth of pulmonary metastases and host
survival, Annals of Surgery 159 (1964) 161-171.
\bibitem{rzeski04}
W. Rzeski, S. Pruskil, A. Macke, U. Felderhoff-Mueser, A.K. Reiher, F. Hoerster,
C. Jansma, B. Jarosz, V. Stefovska, P. Bittigau, C. Ikonomidou, Anticancer
agents are potent neurotoxins in vitro and in vivo, Annals of Neurology 56
(2004) 351-360.
\bibitem{stupp05}
R. Stupp, W.P. Mason, M.J. Van den Bent, M. Weller, B. Fisher, M.J.B. Taphoorn,
K. Belanger, A.A. Brandes, C. Marosi, U. Bogdahn, J. Curschmann, R.C. Janzer,
S.K. Ludwin, T. Gorlia, A. Allgeier, D. Lacombe, J.G. Cairncross, E.
Eisenhauer, R.O. Mirimanoff, Radiotherapy plus concomitant and adjuvant
temozolomide for glioblastoma, The New England Journal of Medicine 352 (2005)
987-996.
\bibitem{strik12}
H.M. Strik, C. Marosi, B. Kaina, B. Neyns, Temozolomide dosing regimens for
glioma patients, Current Neurology and Neuroscience Reports 12 (2012) 286-293.
\bibitem{azevedo09} 
F.A.C. Azevedo, L.R.B. Carvalho, L.T. Grinberg, J.M. Farfel, R.E.L. Ferretti,
R.E.P. Leite, W.J. Filhos, R. Lent, S. Herculano-Houzel, Equal numbers of
neuronal and nonneuronal cells make the human brain an isometrically scaled-up
primate brain, The Journal of Comparative Neurology 513 (2009) 532-541.
\bibitem{vogelzang84}
N.J. Vogelzang, Continuous infusion chemotherapy: A critical review, Journal of
Clinical Oncology 2 (1984) 1289-1304.
\bibitem{rotman91}
M. Rotman, C.J. Rosenthal, Concomitant continuous infusion chemo\-therapy and
radiation, Berlin, Springer-Verlag (1991).
\bibitem{lai17}
J. Lai, P. Xu, X. Jiang, S. Zhou, A. Liu, Successful treatment with
anti-programmed-death-1 antibody in a relapsed natural killer/T-cell lymphoma
patient with multi-line resistance: a case report, BMC Cancer 17 (2017) 507.
\bibitem{beer04}
T.M. Beer, M. Garzotto, W.D. Henner, K.M. Eilers, E.M. Wersinger, Multiple
cycles of intermittent chemotherapy in metastatic androgen-independent prostate
cancer, British Journal of Cancer 91 (2004) 1425-1427.
\bibitem{foo09}
J. Foo, F. Michor, Evolution of resistance to targeted anti-cancer therapies
during continuous and pulsed administration strategies, PLOS Computational
Biology 5 (2009) e1000557.
\bibitem{ren17}
H.-P. Ren, Y. Yang, M.S. Baptista, C. Grebogi, Tumour chemotherapy strategy
based on impulse control theory, Philosophical Transactions of the Royal
Society A 375 (2017) 20160221.
\bibitem{wasan14}
H. Wasan, A.M. Meade, R. Adams, R. Wilson, C. Pugh, D. Fisher, B. Sydes, A.
Madi, B. Sizer, C. Lowdell, G. Middleton, R. Butler, R. Kaplan, T. Maughan,
Intermittent chemotherapy plus either intermittent or continuous cetuximab for
first-line treatment of patients with KRAS wild-type advanced colorectal cancer
(COIN-B): A randomised phase 2 trial, The Lancet Oncology 14 (2014) 631-639.
\end{thebibliography}
\end{document}